# Human Activity Recognition using Multi-Head CNN followed by LSTM

Waqar Ahmad[1], Bibi Misbah Kazmi[1], Hazrat Ali[1]

[1]Department of Electrical and Computer Engineering,
COMSATS University Islamabad, Abbottabad Campus, Abbottabad, Pakistan.
waqarahmad3411@gmail.com, misbahkazmi3@gmail.com, hazratali@cuiatd.edu.pk

*Abstract*— This study presents a novel method to recognize human physical activities using CNN followed by LSTM. Achieving high accuracy by traditional machine learning algorithms, (such as SVM, KNN and random forest method) is a challenging task because the data acquired from the wearable sensors like accelerometer and gyroscope is a time-series data. So, to achieve high accuracy, we propose a multi-head CNN model comprising of three CNNs to extract features for the data acquired from different sensors and all three CNNs are then merged, which are followed by an LSTM layer and a dense layer. The configuration of all three CNNs is kept the same so that the same number of features are obtained for every input to CNN. By using the proposed method, we achieve state-of-the-art accuracy, which is comparable to traditional machine learning algorithms and other deep neural network algorithms.

## I. INTRODUCTION

Human physical activity is the state of the body such as sleeping, walking, laying, eating, jogging and standing. Human activity recognition task has greatly attracted the machine learning research community as it has a revolutionary potential in remote health monitoring, video surveillance, military defense, smart homes, personal fitness, and many more fields.

Human Activity recognition approaches can be categorized broadly into two ways that are sensor-based recognition methods and visual-based recognition methods. In visual-based recognition, one has to record and process visual data in the form of images/videos and then perform recognition with computer vision techniques [1], [2], [3]. Sensor-based recognition methods work on the principle of processing of data recorded by wearable sensors. For example, to monitor elderly or disabled individuals under scenarios where immediate help from a caretaker is not always viable, wearable sensors embedded in sports bracelets or smart watches can provide useful data on the activities of such individuals. The data acquired from wearable sensors is typically useful to monitor the condition of such individuals and to determine if an immediate remedy is required.

In sensor-based recognition, activity recognition is performed using data from wearable sensors (such as those embedded in smartphones, sports bracelets, and smart watches). The latter is used in this work.

For the recognition of activities, feature extraction is the main step to differentiate between different activities by capturing relevant information. The approaches used for the activity recognition depend on the features extracted from the sensors' data like accelerometer and gyroscope, etc.

Smartphone development having accelerometers gives researchers the ability to recognize human physical activities accurately and to achieve a better understanding of the relationship between health and physical activities.

In this study, we use data acquired from the accelerometer and gyroscope embedded in smartphones to extract features for activity recognition. We then propose a multi-head Convolutional Neural Network (CNN) and a long-short term memory network (LSTM) in order to perform the human activity recognition task. The rest of the paper is organized as follows: In Section II, we briefly discuss some of the existing methods used for activity recognition from sensors data. In Section III, we describe the data used in this work. We then explain the proposed model in Section IV. We then report and compare the results in Section V. Finally, the paper is concluded in Section VI.

## II. LITERATURE REVIEW

The traditional physical activity recognition algorithms showed low accuracy due to variable, complex and dynamic features. Wang *et al.* [4] proposed a method based on LSTM named as hierarchical deep LSTM (H-LSTM). Starting from preprocessing to smooth and de-noising the original data from the sensor and then using the time-frequency-domain method, the authors in [4] selected and extracted the features. Then for the classification of the activities, the authors used H-LSTM. So, by using three UCI datasets, the authors in [4] conducted an experiment on the extraction of features' vectors automatically and classified the human physical activities. Tamamori *et al.* [5] developed a life-logging system using smartphone sensors. RNN and feed-forward neural network (FFNN) were considered as the effective classifiers for the human activity recognition (HAR) task. The authors in [5] conducted an experiment to record the data by building a life-logging prototype system that included both indoor and outdoor activities. The shortcoming of the proposed method is that the RNN has a problem of gradient vanishing so the accuracy can be low using the RNN with an FFNN.

Suto *et al.* [6] investigated the performance of the different artificial neural networks (ANN) architectures on two publicly available data sets. The results showed that the preprocessing of data and hyper-parameter settings are the key factors in ANN because the difference between the accuracy of well parameterized and a poorly parameterized ANN is large. A well-tuned ANN performs better than other traditional machine learning methods in human activity recognition (HAR). The authors in [6] reported good accuracy scores, but the execution time of the proposed method is very high, which cannot be used in real-world scenarios. Lee *et al.* [7] proposed a method based on one-dimensional CNN for recognition of human activities using tri-axial accelerometer data collected through smartphones. The tri-axial acceleration data was transformed into the vector magnitude data and was used as the input to the one-dimensional CNN. The authors in [7] used a very simple method, but the accuracy claimed by the author is low as compared to other one-dimensional CNN. Ronao *et al.* [8] proposed a deep CNN for the human activity recognition task. The authors used raw readings of the sensors for the training of the model. The experiments performed showed that with the addition of every layer, CNN derived more complex features, and the complexity difference level decreases with the addition of every new layer.

Lu *et al.* [9] employed an unsupervised method for the recognition of human physical activities using smartphone accelerometers. Extraction of features was done from the raw acceleration data that was collected from smartphones, then for the activity recognition, a method called MCODE was used. Lui *et al.* [10] presented an algorithm that identified the temporal patterns and utilized those patterns for the automated recognition of activities. Ha, *et al.* [11] presented CNN-pf and CNN-pff for multi-modal data. The authors in [11] employed partial weight sharing and full weight sharing both for CNN models so that the common characteristics and modality-specific characteristics were learnt from multi-modal data and in upper layers, data was aggregated. The aforementioned models are based on the idea of training a single neural network (or a CNN) for the different types of sensor data. In comparison, we propose to use a CNN ensemble architecture comprising of three CNNs working in parallel. While each CNN is trained independently, their combined working strategy results in better performance as compared to a stand-alone CNN model.

## III. DATASET

The dataset used for the proposed architecture in this paper is "UCI human activity recognition using smartphone dataset" [14]. This dataset was built from the data acquisition from 30 people with age in the range of 19 to 48 years. The individuals were doing daily life work carrying the smartphone at waist position and the smartphone was embedded with the inertial sensors. There are six actions performed by each person i.e. "*Walking, Walking Upstairs, Walking Downstairs, Sitting, Standing, Laying*". The data in the dataset is already processed by noise filters and sampled using a 2.5 seconds window with an overlap of 50%. The data is acquired using tri-axial linear acceleration and tri-axial gyroscope at a rate of 50Hz. The acceleration sensor captures a signal having gravitational and body motion components. These components are separated from each other using a Butterworth filter with 0.3 Hz cutoff frequency. So, in the dataset, we see three streams of data. i.e. x y and z-axis of total acceleration, body acceleration, and body gyroscope. 70% of training data is used for training and 30% of data is used for testing.

In order to evaluate our model on the data we first prepare our data. We did the standardization of the dataset. After applying the normalization, the dataset has zero mean and unit variance. But this transformation only makes sense if the distribution among the data is Gaussian distribution. To check the distribution among the data, we plot the data set. The plots are shown in Fig. 1, which reveals a Gaussian distribution for the data.

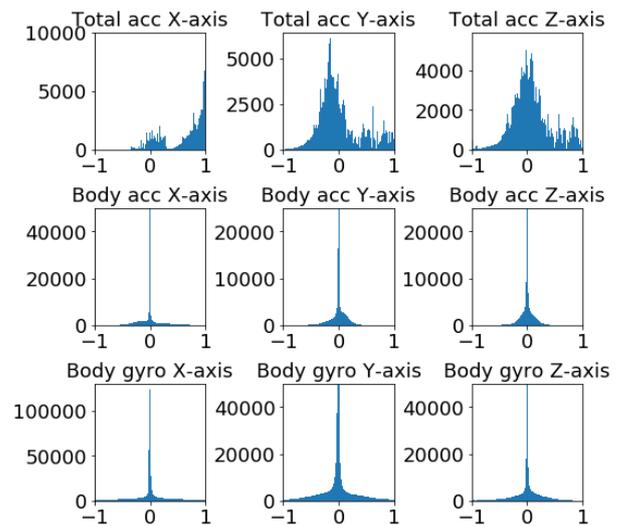

Fig. 1 Gaussian distribution plot for the data

## IV. PROPOSED METHODOLOGY

In this work, a multi-head CNN architecture is used, which is followed by the LSTM model to perform the recognition task of human physical activities. Conventionally, CNN was designed for images and involved convolution of the input with a 2D filter mask. However, the sensor data used in this work is a time series data. Hence, the filter mask involved is one-dimensional, which is convoluted with the one-dimensional time-series data to extract useful features. Recurrent Neural Network (RNN) are typically popular with time-series data. However, RNN may suffer the gradient vanishing problem and thus, LSTM is used in this work.

LSTM network has memory cells, which remember sequence information for a very long time. LSTM has a memory block with recurrent connection to itself and three multiplicative gates that are input, forget and output gate, as shown in Fig. 2.

Forget gate decides which information is to be eliminated from the block on the basis of a given condition. The input gate decides to update the state of memory for a given input. The output gate decides what the output is to be given an input and a memory state. For more information on LSTM, please refer to [13].

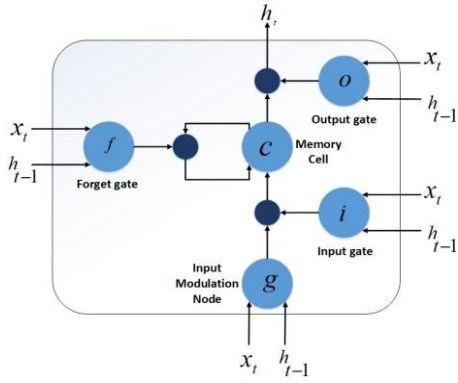

Fig. 2 A memory block of LSTM

The CNN architecture used in this paper has four convolutional layers, each followed by a max-pooling layer. Each convolutional layer has a filter size of 3 and stride of 1, the values been selected after a grid search. So, we trained our model on a filter of size three. To increase the accuracy of recognition of human physical activity, we propose the multi-head CNN architecture comprising of three parallel CNN models. As shown in Fig. 4, the three CNNs are implemented in the proposed model, which are followed by the LSTM layers, a dense layer, and an output layer. All the three parallel CNNs used in the proposed method have the same configuration. The configuration of a single CNN is shown in Table I. The parameter used while training of the proposed model is shown in Table II.

### A. Single CNN followed by LSTM:

The CNN configuration is shown in Table I is used in the model which is followed by LSTM layers. Each LSTM has 128 units. First, we trained a model with a single CNN followed LSTM layer and a fully connected layer. The inputs to the model are the nine values representing the total acceleration, body acceleration, and the body gyroscope, respectively. All the input in this step are provided to the one-dimensional CNN to extract some useful features. The features thus extracted are then fed to the LSTM layer having 128 units, which is followed by a fully connected layer having 1000 neurons and an output layer having 6 neurons, as shown in Fig. 3. By using this architecture for human physical activity recognition, we got a very good accuracy with as presented in the results section.

### B. Multi-head CNN followed by LSTM

We trained a multi-head CNN architecture by using three CNN architectures connected in parallel, as shown in Fig. 4. In which the Single one-dimensional CNN discussed above is used in parallel. And the output of all parallel CNNs is merged and provided as input to the LSTM layer, which is followed by a fully connected layer having 1000 neurons and an output layer having neurons equal to number is outputs i.e. 6 neurons. In order to use the proposed multi-head CNN, we split our input into three streams. 1) total acceleration. 2) body acceleration and 3) body gyroscope. At the input of the first CNN, x, y and z-axis of total acceleration are used as input. The input to first CNN is (1 x 3), i.e. 1 row and 3 columns, While the input of the second CNN is the input of body acceleration data having the dimension of 1x3. Same for the third CNN which has input of body gyroscope having dimension 1×3. Each CNN has the same configuration, so each CNN extracts features for its corresponding inputs. Following this, the output sequences are merged and forwarded to the LSTM layer. The LSTM model has 128 units. The output of the LSTM model is forwarded to a fully connected layer of 1000 neurons. This is followed by an output layer, which has only six neurons on the basis of the number of output classes. A 'softmax' classifier is introduced to classify the output of final layer on the basis of the probabilities generated by the output layer. The confusion matrices of single CNN-LSTM architecture and multi-head CNN-LSTM architecture is shown in Fig. 10 and Fig. 11 respectively in the result section.

TABLE I. CONFIGURATION OF SINGLE HEAD 1D CNN

| Layers | Parameter |
|---|---|
| Softmax | No. of classes = 6 |
| Fully connected | No. of neurons = 6 |
| Fully connected | No. of neurons = 1000 |
| Dropout | Prop: 0.3 |
| LSTM | 128 units |
| Max pooling | Filter Size = 2, Stride = 1 |
| Convolution 1D | No. of filters = 32, Size of filter = 3 , Stride = 1 |
| Max pooling | Size of filters = 2, Stride = 1 |
| Convolution 1D | No. of filters = 64, Size of filter = 3, Stride = 1 |
| Max pooling | Size of filter = 2, Stride = 1 |
| Convolution 1D | No. of filters = 128, size of filter = 3, Stride = 1 |
| Max pooling | Size of filters = 2, Stride = 1 |
| Convolution 1D | No. of filters = 512, size of filter = 3, Stride = 1 |
| Input | $1 \times 9$ |

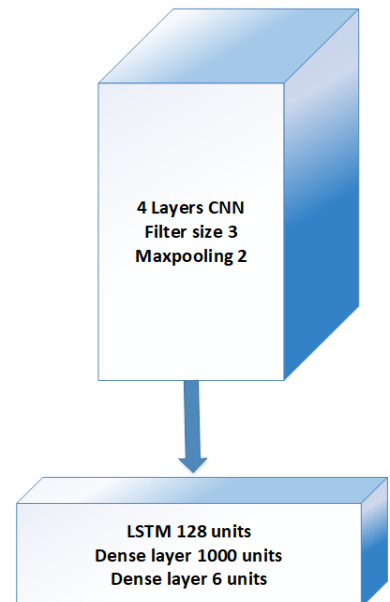

Fig. 3 Single CNN-LSTM Architecture

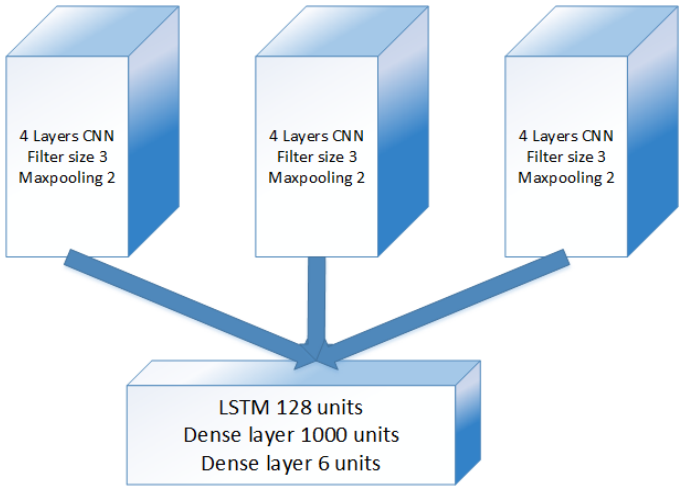

Fig. 4 Multi-Head CNN-LSTM Architecture

TABLE II. CONFIGURATION OF SINGLE HEAD 1D CNN

| Layer Type | Parameter |
| --- | --- |
| Optimizer | Adam |
| Learning Rate | 0.001 |
| Beta_1 | 0.9 |
| Beta_2 | 0.999 |
| Loss Function | Categorical Cross Entropy |
| Batch Size | 32 |
| Epochs | 17 |

### C. Traditional Machine Learning Methods

In order to have a comparison with traditional machine learning models, we also train and evaluate SVM and KNN on the same dataset. For the training of these models, we take advantage of features as provided with the dataset. There is a total of 582 features provided for the dataset. These features include standard deviation, kurtosis, skewness, and root mean square values besides many other parameters. The aforementioned feature along with some FFT features are calculated and a feature vector is provided. By using this feature vector, we train an SVM model with polynomial kernel and we train a KNN classifier. For both the SVM and KNN classifiers, we divide the data into 70% training data and 30% testing data. We evaluate KNN for different values of K. We achieve the best accuracy score for K=10. The graph showing the error values for the KNN classifier with different values of K is shown in Fig. 7. While processing SVM and KNN for the UCI HAR dataset, we noticed that KNN is relatively slower as compared to SVM. Naturally, if we further increase the dataset size, this may cause the slowing down of the KNN model. The confusion matrix for SVM and KNN is shown in Fig. 5 and Fig. 6 respectively.

## V. EXPERIMENTAL RESULTS AND ANALYSIS

### A. Evaluation Criteria:

We evaluate our proposed method in terms of precision, recall, and F1-score. Table III shows the scores for the Polynomial SVM, KNN, single CNN-LSTM, and multi-head CNN-LSTM architectures. Table IV shows the accuracy of single CNN-LSTM and multi-head CNN-LSTM architectures. The use of one-dimensional CNN for time-series signals has recently been popular within the research community and a few attempts have been reported in [7], [12] and [8]. We compare our results with these methods as reported in Table V.

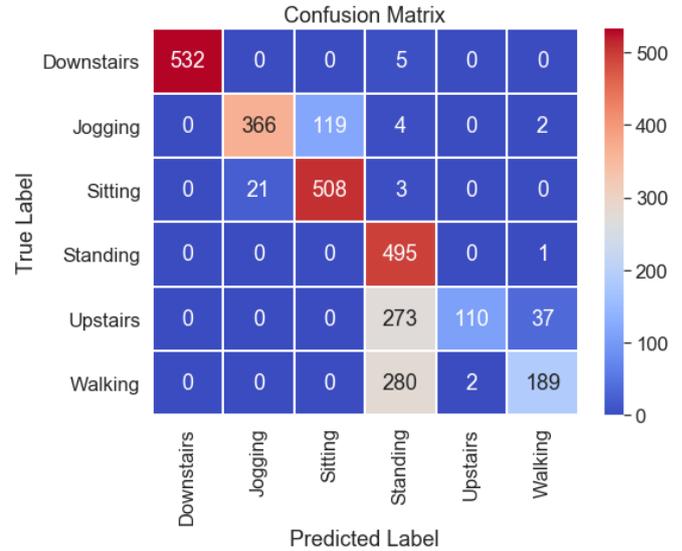

Fig. 5 Confusion Matrix for SVM

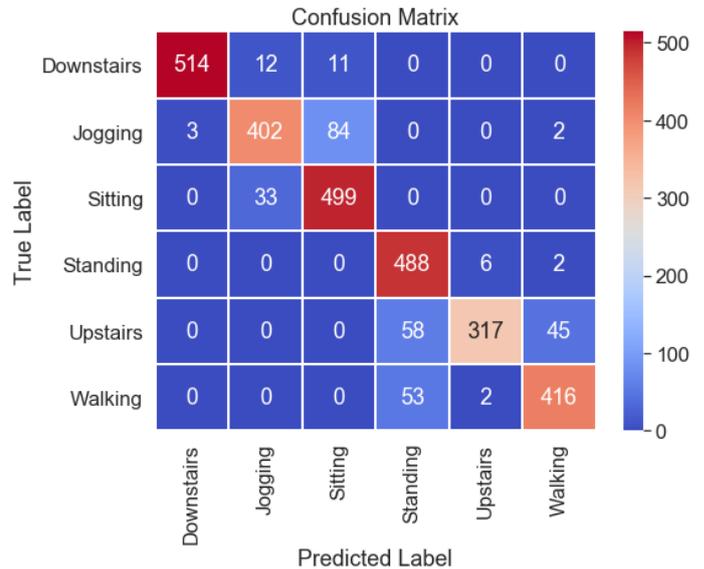

Fig. 6 Confusion Matrix for K-Nearest neighbor

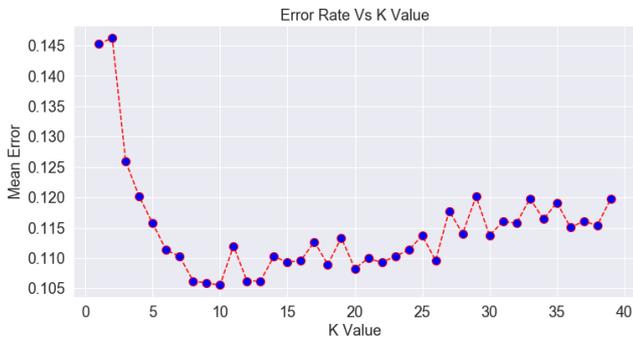

Fig. 7 Error vs K for the KNN classifier

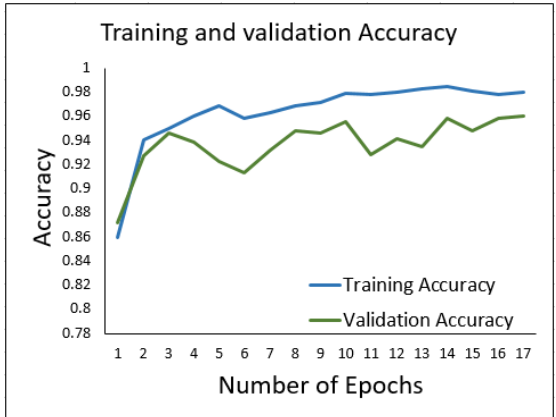

Fig. 8 Accuracy Graph for Multi-head CNN-LSTM

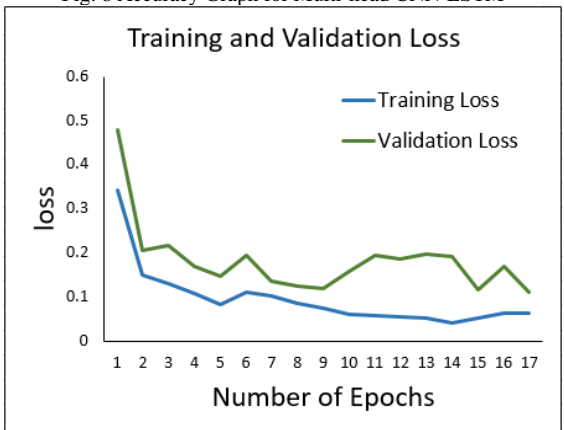

Fig. 9 Loss graph of Multi-head CNN-LSTM

### B. Training of CNN-LSTM Model

The experiment of training the single CNN followed by LSTM is done on Core i3 CPU having a 2.8 GHz processor and 4 GB of RAM. The training time for the proposed model was 8 hours on the specified hardware. The proposed model is trained in an end-to-end method i.e. the parallel CNN along with the LSTM is trained at the same time. The model is tested for 50 epochs, 30 epochs, and 20 epochs. While using 50 epochs with a batch size of 32 the model diverges beyond 17 epochs. The reason behind the divergence of the model is that we are using a fixed learning rate as specified in table II. Hence, we select 17 epochs with a batch size of 32. The accuracy values obtained for the single CNN followed by LSTM and the multi-head CNN followed by LSTM models over 17 epochs are reported in Table IV. The loss and accuracy graph throughout the 17 epochs are shown in Fig. 8 and Fig. 9 respectively.

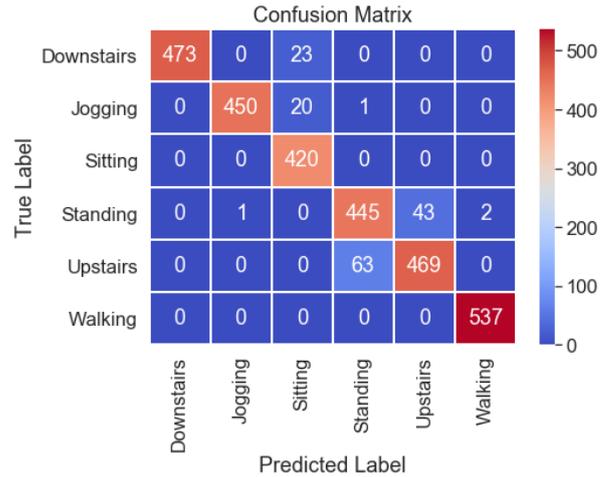

Fig. 10 Confusion Matrix for Single CNN-LSTM

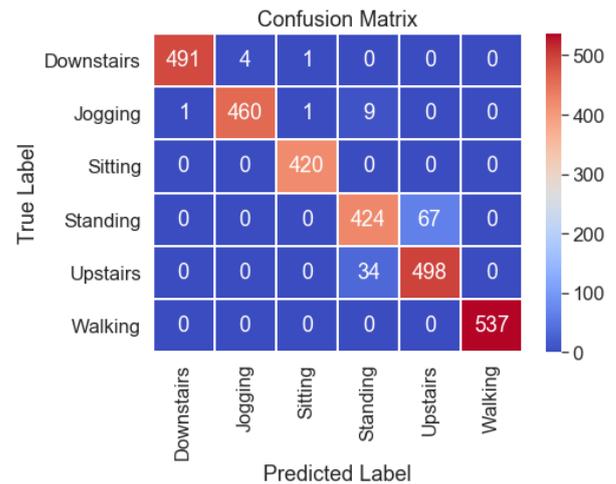

Fig. 11 Confusion Matrix for Multi-head CNN-LSTM

TABLE III. PERFORMANCE COMPARISON

| Algorithm Names | Precision | Recall | F1-score | Support |
|---|---|---|---|---|
| Polynomial SVM | 0.84 | 0.75 | 0.73 | 2947 |
| KNN | 0.90 | 0.89 | 0.89 | 2947 |
| Single CNN-LSTM | 0.95 | 0.95 | 0.95 | 2947 |
| **Multi-head CNN-LSTM** | **0.96** | **0.96** | **0.96** | **2947** |

TABLE IV. ACCURACY SCORES

| Algorithm | Accuracy |
|---|---|
| Single CNN-LSTM | 94.1 % |
| **Multi-head CNN-LSTM** | **95.76 %** |

### C. Analysis on the Results:

The comparison of the accuracy of our proposed model with other neural network models is shown in table V. Form the table V it is very clear that by using the proposed model with the raw data as input we can achieve a state-of-the-art accuracy and there is no need of manually calculating other features like frequency domain feature that is used by the authors in [8]. While comparing the accuracy score of the single stream of the proposed model i.e. single one-dimensional CNN followed by LSTM with the other one-dimensional CNN [7], it is clear that out proposed single CNN followed by LSTM perform better and report a fair increase inaccuracy. From the accuracies reported in table III, we conclude that our proposed model performs better than traditional machine learning methods that are already used for human physical activity recognition. The main reason behind achieving better accuracy with the proposed model is that we split the data stream into 3 parallel streams and used three one-dimensional CNNs in parallel to extract the features for each stream separately, which is than merged together before feeding into LSTM layers. The significance of the proposed model is that instead of mapping one-dimensional data to two-dimensional data and then using two-dimensional CNN which is computationally expensive, we just used one-dimensional CNN which is followed by LSTM layers, as the data obtained from the sensors is also one-dimensional. so practically the proposed model is simple and computationally cheap.

TABLE V. PERFORMANCE COMPARISON

| Method | Method | Accuracy | Year |
|---|---|---|---|
| Ronao [8] | Convnet+MLP | 94.79% | 2016 |
| Ronao [8] | FFT+Convnet | 95.75% | 2016 |
| Lee [7] | 1D CNN | 92.71% | 2017 |
| Jiang [12] | DCNN | 95.18% | 2015 |
| **Proposed** | **Multi-head CNN+LSTM (Raw data)** | **95.76%** | - |

VI. CONCLUSION AND FUTURE WORK

In this work, we proposed a novel multi-head CNN followed by LSTM architecture to recognize human physical activity recognition. We used the UCI database in which the data is divided into training and test subsets with a 7:3 ratio respectively. The data is collected for six different activities like sitting, standing, walking, walking upstairs, walking downstairs and laying. More specifically, we compared our results with both traditional machine learning methods such as SVM as well as more recent deep learning methods used for human activity recognition. Experimental results on the UCI dataset show that the proposed multi-head CNN-LSTM approach is promising in terms of the performance accuracy when compared with many other methods. The proposed multi-head CNN architecture resulted in better accuracy compared to a single CNN model as well as SVM and KNN classifiers for the human activity recognition task.